\documentclass[UTF-8,preprint,showpacs,preprintnumbers,superscriptaddress]{revtex4-1}

\usepackage{amsmath,amssymb,graphicx,bm,subfigure,placeins,mathrsfs}
\usepackage[colorlinks=true,          
            linkcolor=red,          
            citecolor=blue,         
            urlcolor=blue]{hyperref} 
            
\begin{document}
	\title{Spherically symmetric charged (anti-)de Sitter black hole in $f(R,T)$ gravity coupled with nonlinear electrodynamics}

\author{Tianyou Ren}
	\affiliation{College of Physics Science and Technology, Hebei University, Baoding 071002, China}	

\author{Zhenglong Ban}
	\affiliation{College of Physics Science and Technology, Hebei University, Baoding 071002, China}

\author{Yaobin Hua}
	\affiliation{College of Physics Science and Technology, Hebei University, Baoding 071002, China}

\author{Rong-Jia Yang \footnote{Corresponding author}}
\email{yangrongjia@tsinghua.org.cn}
\affiliation{College of Physics Science and Technology, Hebei University, Baoding 071002, China}
\affiliation{Hebei Key Lab of Optic-Electronic Information and Materials, Hebei University, Baoding 071002, China}
\affiliation{National-Local Joint Engineering Laboratory of New Energy Photoelectric Devices, Hebei University, Baoding 071002, China}
\affiliation{Key Laboratory of High-pricision Computation and Application of Quantum Field Theory of Hebei Province, Hebei University, Baoding 071002, China}

\begin{abstract}
We investigate black hole solutions in quadratic $f(R,T)=R+kT^2$ gravity coupled with nonlinear electrodynamics $\mathcal{L}(F)=\alpha - F/(4\pi) + \gamma F^2$. Solving the field equations yields an exact magnetically charged (anti-)de Sitter black hole metric featuring novel $r^{-6}$ and $r^{-14}$ correction terms. The horizon structure can exhibit up to four horizons depending on the parameters. Thermodynamic analysis reveals that the heat capacity exhibits critical phase transitions, with the critical point significantly shifted by the coupled corrections. Using the effective metric formalism, we study the photon sphere and black hole shadow, finding that the magnetic charge $Q$ shrinks the shadow while the nonlinear parameter $\gamma$ enlarges it. Constraints from Event Horizon Telescope observations of M87* and Sgr A* place the parameter $\gamma$ at the order of $10^3$ at $1\sigma$ and $2\sigma$ confidence levels. Our results demonstrate that quadratic $f(R,T)$ gravity combined with nonlinear electrodynamics provides a consistent framework for strong-field gravity phenomenology.

\end{abstract}

\keywords{modified gravity, $f(R,T)$ gravity, nonlinear electrodynamics, black hole shadows}

\maketitle

\section{Introduction}
\label{1}
Since the establishment of General Relativity (GR) \cite{Einstein:1916vd}, it has demonstrated strong applicability in explaining the precession of Mercury's perihelion, gravitational lensing effects, recent observations of gravitational waves, and black hole (BH) imaging. However, it exhibits limitations in addressing issues such as dark-matter distribution, late-time cosmic accelerated expansion, and quantum effects in strong gravitational fields. This has spurred the development of modified gravity theories. One of the most promising avenues is $f(R)$ gravity \cite{DeFelice:2011jm,Nojiri:2006ri,Sotiriou:2008rp}, where the Einstein-Hilbert Lagrangian is replaced by an arbitrary function of the Ricci scalar $R$. However, although viable models do exist, many $f(R)$ models face stringent constraints from local gravitational tests and solar system observations. For further generalization, theories involving the coupling between geometry and matter were developed, leading to modified gravity theories where the gravitational Lagrangian is a function of $R$ and the matter Lagrangian $\mathcal{L_{\text{mat}}}$ \cite{Jaybhaye:2022gxq,Kavya:2022dam,Garg:2023cfn,Shukla:2023xlz,VijayaSanthi:2025tjf}, or a function of $R$ and the trace $T$ of the energy-momentum tensor, thus giving rise to the so-called $f(R,T)$ gravity \cite{Harko:2011kv,Fisher:2019ekh, Yang:2015jla}, which breaks through the GR framework where curvature is only linearly determined by matter, by introducing a coupling between the Ricci scalar $R$ and the trace $T$ of the energy-momentum tensor, becoming an important tool for investigating gravity-matter interactions.
The $f(R, T)$ theories have been extensively validated against observational data, including Type Ia supernovae \cite{Siggia:2024pgr,SupernovaCosmologyProject:1997zqe,SupernovaCosmologyProject:1998vns} and cosmic microwave background radiation \cite{Rudra:2020nxk}. Its applications span diverse areas of gravitational physics: in cosmology, it provides mechanisms for explaining early universe evolution \cite{Jamil:2011ptc,Shabani:2013djy,Shabani:2014xvi,Myrzakulov:2012qp,Yang:2015jla}; in stellar physics, it offers insights into compact objects such as white dwarfs, exotic stars \cite{Carvalho:2017pgk,Bhattacharjee:2023olf,Deb:2018gzt,Moraes:2015uxq,Maurya:2019hds,Maurya:2019sfm,Feng:2022bvk}, and gravastars \cite{Das:2017rhi,Yousaf:2019zcb}; in wormhole physics, it enables the construction of novel solutions \cite{Azizi:2012yv,Zubair:2016cde,Moraes:2016akv,Moraes:2017mir,Elizalde:2018frj}; and in foundational studies, it facilitates investigations of energy conditions \cite{Sharif:2012gz,Sharif:2012ce,Alvarenga:2012bt} and thermodynamic properties \cite{Sharif:2012zzd}. Furthermore, in quantum cosmology, $f(R, T)$ gravity has been explored to understand quantum behavior under the Friedmann-Robertson-Walker metric \cite{Xu:2016rdf}. These diverse applications demonstrate how $f(R, T)$ gravity not only extends the scope of GR, but also provides promising pathways to address some of the most profound unresolved questions.

In this work, we extend the matter-geometry coupling in $f(R,T)$ gravity by incorporating a quadratic term in the trace of the energy-momentum tensor, i.e., $f(R,T)=R + k T^2$, where $k$ is a coupling constant. This choice is motivated by the desire to explore higher-order backreaction effects of matter fields on spacetime curvature. Unlike the widely adopted linear coupling form (e.g., $f(R,T)=R+kT$) \cite{Rois:2025tfe}, the $T^2$ term introduces a nonlinear response of the energy-momentum tensor, significantly altering the weight of the matter term in the field equations. It is important to note that in $f(R,T)$ theories, the conservation of the energy-momentum tensor is generally not guaranteed, i.e., $\nabla^\mu T_{\mu\nu} \neq 0$ \cite{Harko:2011kv}. This non-conservation is a physical feature rather than a flaw, as it implies an energy transfer between matter and geometry, which can be interpreted as the effect of the curvature-matter coupling. In our specific model, the $T^2$ term will regulate the higher-order corrections to the metric functions in a more intricate manner compared to its linear counterpart, potentially leading to novel predictions in strong-field regimes.

Meanwhile, nonlinear electrodynamics (NLED) has been extensively studied as a method to resolve classical singularities inherent in Maxwell's theory and to explore strong electromagnetic field effects in the early universe and compact objects. NLED models incorporate higher-order terms into the electromagnetic Lagrangian, which become significant in high-field-strength regimes. Previous studies within the $f(R,T)$ framework have often employed linear coupling forms combined with power-law NLED \cite{Rois:2025tfe}, successfully deriving static spherically symmetric charged BH solutions with an effective cosmological constant. Other studies have further extended NLED to the Euler-Heisenberg (EH) model \cite{Liang:2025hzr,MosqueraCuesta:2004wh}, revealing the multi-horizon structure of BHs and the classification of photon orbits under quantum vacuum polarization effects. While the existing $f(R,T)$-NLED framework covers linear coupling and various NLED forms, there is room for expansion in two key aspects. First, the coupling form of $f(R,T)$ has mostly been limited to the first-order term of $T$, while the higher-order $T^2$ coupling we propose introduces a richer matter-geometry interaction, potentially affecting spacetime singularity and horizon stability in strong gravitational fields. Second, the choice of the NLED Lagrangian needs to balance reasonable asymptotic behavior with solvability. To this end, we consider a simplified NLED Lagrangian $\mathcal{L}(F) = \alpha - \frac{1}{4\pi}F + \gamma F^2$ within the quadratic $f(R,T)$ gravity framework. Here, $\alpha$ acts as the source of an effective cosmological constant, and $\gamma$ characterizes the strong-field self-interaction of the electromagnetic field. This form can revert to standard Maxwell theory in weak fields while introducing nonlinear corrections via the $\gamma F^2$ term, which helps to regulate the central singularity and modify the horizon structure.

Recent investigations have explored various aspects of BH physics in modified gravity and nonlinear electrodynamics, including the shadow and parameter constraints of rotating black holes in different theoretical frameworks \cite{Raza:2023vkn,Raza:2025ohk,Zubair:2022fiw}. Since the discovery of black hole thermodynamics \cite{Bekenstein:1973ur, Hawking:1975vcx}, extensive studies have been devoted to understanding the thermal properties of black holes, including phase transitions and stability analyses in various gravity theories \cite{Chaudhary:2024lzm}. Our work contributes to this line of inquiry by introducing a novel coupling mechanism—quadratic $f(R,T)$ gravity combined with a quadratic NLED Lagrangian—and exploring its implications for both the interior (singularity/horizon) and exterior (photon orbit/shadow) structure of BHs. The metric function we derive from this model contains novel correction terms with clear physical significance, including a $r^{-6}$ term originating from the interplay between $T^2$ gravity and NLED, and a $r^{-14}$ term from higher-order nonlinear effects. These terms provide a concise yet complete theoretical framework for systematically analyzing BH horizon distribution, thermodynamic properties, and photon orbit dynamics. Compared to the linear $f(R,T)=R+kT$ model, our quadratic coupling yields a more complex interplay between matter and geometry, leading to corrections at different orders that become significant at distinct scales.

The field of BH astrophysics has been revolutionized by the Event Horizon Telescope (EHT) collaboration. Using Very-Long-Baseline Interferometry (VLBI), the EHT team captured the first direct image of the supermassive BH at the center of galaxy M87 \cite{EventHorizonTelescope:2019dse}. Soon after, they imaged Sagittarius A* (Sgr A*), the supermassive BH at our Milky Way's center \cite{EventHorizonTelescope:2022wkp}. Both images show similar ring-like structures with central dark areas surrounded by bright emission rings. These discoveries provide new ways to study extreme gravitational environments. The theoretical interpretation of these shadows is rooted in the strong gravitational lensing of light. Early studies by Synge \cite{Synge:1966okc} and Bardeen \cite{Bardeen:1972fi} established the foundation for calculating photon orbits and shadows in Schwarzschild and Kerr spacetimes. Building on early theoretical works, subsequent studies have further extended the calculations of black hole shadows to more sophisticated gravitational theories and carried out parameter constraints in combination with EHT observational data. For example, the shadow and accretion disk of the Kerr-Sen black hole in Einstein-Maxwell-dilaton-axion gravity have been studied and constrained by EHT observations \cite{Feng:2024iqj}. Similar analyses have also been performed for black holes in quantum fluctuation modified gravity \cite{Yin:2025coq}. Beyond null geodesics, timelike periodic orbits in quantum-corrected black holes have also been investigated for their gravitational wave signatures \cite{Yang:2024lmj}. Building on these advances, we study how modified gravity combined with nonlinear electrodynamics affects photon orbits and shadow properties. In particular, we will use the effective metric formalism \cite{Novello:1999pg,Toshmatov:2021fgm} to account for the nonlinear electromagnetic effects on photon propagation, and then confront our theoretical predictions with the EHT observations of M87* and Sgr A* to constrain the model parameters. This allows us to assess the viability of our quadratic $f(R,T)$ model in the strong-field regime.

The structure of this paper is as follows. In Section. \ref{2}, we derive the field equations for the coupling between $f(R,T)$ gravity and NLED. In Section. \ref{3}, we derive the BH solution for the model under consideration and analyze its horizon structure. In Section. \ref{4}, we investigate the thermodynamic properties, including temperature, entropy, free energy, and heat capacity, and discuss the stability and phase transitions of the black holes. In Section. \ref{5}, we analyze the effective potential and the photon geodesic structure, compute the BH shadow, and constrain the model parameters using EHT data. In Section. \ref{6}, we summarize the research findings and outline future directions for extension. To simplify the discussion, we use natural units ($G=c=1$) throughout the article.

\section{field equations}
\label{2}
The theoretical framework is constructed by the combination of $f(R,T)$ gravity and a NLED matter term, with the total action defined as
\begin{equation}
S = \int \sqrt{-g} \left[ \frac{1}{16\pi} f(R,T) + \mathcal{L}(F) \right] d^4x,
\end{equation}
where $g$ is the determinant of the metric $g^{\mu\nu}$, $R$ is the Ricci scalar, $T$ denotes the trace of the energy-momentum tensor and $F= \frac{1}{4}F^{\mu\nu}F_{\mu\nu}$. The gravitational part adopts a quadratic coupling form
\begin{equation}
\label{eq:f(R,T)}
f(R,T)=R+kT^2,
\end{equation}
The matter part employs a quadratic NLED Lagrangian
\begin{equation}
\label{eq:L(F)}
\mathcal{L}(F)=\alpha-\frac{1}{4\pi} F+\gamma F^2,
\end{equation}
%Varying the action with respect to the metric $g^{\mu\nu}$ and the electromagnetic potential $A_\mu$, respectively, yields the complete set of field equations

To obtain the electromagnetic field equations, we vary the action with respect to $A_\mu$, yielding the following equations
  \begin{equation}
  \label{eq:maxwell_modified} 
\nabla_{\mu} \left[ (2f_T(R,T) \, \mathcal{L}_{FF} F - 8\pi \mathcal{L}_F) \, F^{\mu\alpha} \right] =0,
\end{equation}
where $f_T=\frac{\partial{f(R,T)}}{\partial{T}}$. On the other hand, varying the action with respect to the metric yields the gravitational field equations
\begin{equation}
\label{eq:gravity_modified} 
f_RR_{\mu\nu}-\frac{1}{2}fg_{\mu\nu}+(g_{\mu\nu}\Box-\nabla_\mu\nabla_\nu)f_R=8\pi T_{\mu\nu}-f_T(T_{\mu\nu}+\Theta_{\mu\nu}),
\end{equation}
where $f_R=\frac{\partial{f(R,T)}}{\partial{R}}$ and $\Box=g^{\mu\nu}\nabla_\mu\nabla_\nu$. Generally, the matter field tensors $T_{\mu\nu}$ and $\Theta_{\mu\nu}$ are defined, respectively, as
\begin{equation}
T_{\mu \nu} = -\frac{2}{\sqrt{-g}} \frac{\delta \mathcal{L}_{\text{mat}} \sqrt{-g}}{\delta g^{\mu \nu}},
\end{equation}
\begin{equation}
\Theta{\mu \nu} = g^{\sigma \rho} \frac{\delta T_{\sigma \rho}}{\delta g^{\mu \nu}}.
\end{equation}
According to these definitions, we provide the expressions for the energy-momentum tensor $T_{\mu\nu}$ and $\Theta_{\mu\nu}$ of the matter considered here
\begin{equation}
T_{\mu\nu} = g_{\mu\nu}\mathcal{L}(F) - \mathcal{L}_F F_{\mu\rho} F^\rho_\nu,
\end{equation}
\begin{equation}
\Theta_{\mu\nu} = -g_{\mu\nu}\mathcal{L}(F) + \mathcal{L}_F F_{\mu\rho} F^\rho_\nu - 2\mathcal{L}_{FF} F F_{\mu\rho} F^\rho_\nu,
\end{equation}
where $\mathcal{L}_F = \frac{\partial \mathcal{L}(F)}{\partial F}$ and $\mathcal{L}_{FF} = \frac{\partial \mathcal{L}_F}{\partial F}$. 

\section{Black hole solutions and horizon structure}
\label{3}
In this section, we will solve the field equations to obtain BH solutions and then analyze the structures of the horizon.

\subsection{Spherically symmetric charged (anti-)de Sitter black holes}
We consider a static spherically symmetric metric field which takes the form
  \begin{equation}
      ds^2 = -A(r) dt^2 +\frac{1}{B(r)}dr^2 + r^2 (d\theta^2 + \sin^2\theta d\phi^2).
  \end{equation}
We here focus on magnetically charged BHs. To obtain such a BH solution, it is necessary to consider a specific electromagnetic field distribution satisfying the modified Maxwell equation \eqref{eq:maxwell_modified}. We take the nonzero components of the electromagnetic tensor $F_{\mu\nu}$ as follows
\begin{equation}
\label{eq:F_mu_nu} 
F_{\theta\phi} = -F_{\phi\theta} = Q \sin\theta,
\end{equation}
which leads to
\begin{equation}
    F = \frac{Q^2}{2r^4},
\end{equation}
where $Q$ represents the magnetic charge. Inserting Eqs. (\ref{eq:L(F)}) and (\ref{eq:f(R,T)}) into the modified gravitational field equations (\ref{eq:gravity_modified}), we can derive three independent components of the gravitational field equations
\begin{equation}
\label{eq:gravity-00} 
\begin{aligned}
& 16 \alpha^2 k r^{16}-8 \alpha \gamma k Q^4 r^8+16 \pi  \alpha r^{16}-2 r^{15} B'(r)-2 r^{14} B(r)\\&+\gamma^2 k Q^8+4 \pi  \gamma Q^4 r^8-2 Q^2 r^{12}+2 r^{14}=0,
  \end{aligned}
\end{equation}
\begin{equation}
\label{eq:gravity-11} 
\begin{aligned}
 \frac{A'(r)}{r A(r)}-\frac{4 \gamma Q^4 r^8 (\pi -2 \alpha k)+2 r^{12} \left(8 \alpha r^4 (\alpha k+\pi )-Q^2+r^2\right)-2 r^{14} B(r)+\gamma^2 k Q^8}{2 r^{16} B(r)}=0,
  \end{aligned}
\end{equation}
\begin{equation}
\label{eq:gravity-22/33} 
\begin{aligned}
   &\frac{6 \gamma Q^4 (\pi -2 \alpha k)}{r^6}-\frac{8 \alpha r^4 (\alpha k+\pi )+Q^2}{r^2}-\frac{r^2 B(r) A'(r)^2}{4 A(r)^2}\\&+\frac{r \left(r A'(r) B'(r)+2 B(r) \left(r A''(r)+A'(r)\right)\right)}{4 A(r)}+\frac{1}{2} r B'(r)+\frac{7 \gamma^2 k Q^8}{2 r^{14}}=0.
    \end{aligned}
\end{equation}
By combining Eqs. (\ref{eq:gravity-00}), (\ref{eq:gravity-11}) and (\ref{eq:gravity-22/33}), we obtain the BH solution as
\begin{equation}
\label{eq:metric}
A(r)=B(r)=1-\frac{2M}{r}+\frac{Q^2}{r^2}+\frac{8}{3} \alpha  (\alpha k+\pi )r^2+\frac{4 \alpha \gamma k Q^4}{5 r^6}-\frac{2 \pi  \gamma Q^4}{5 r^6}-\frac{\gamma^2 k Q^8}{26 r^{14}}.
\end{equation}
We observe that the $r^2$ term appearing in the solution may be induced by the correction terms of $f(R,T)$ gravity and the $\alpha$ term in the NLED Lagrangian as an effective cosmological constant $\Lambda_{\mathrm{eff}}$. The correction term $r^{-6}$ consists of two parts: $\frac{4 \alpha \gamma k Q^4}{5 r^6}$ originates from the coupling between $f(R,T)$ and NLED, while $-\frac{2 \pi \gamma Q^4}{5 r^6}$ is purely due to the contribution of NLED. The $r^{-14}$ term results from higher-order nonlinear effects coupled with gravity and becomes significant only at extremely small scales.

At the same time, we note that when taking $\alpha=0, ~\gamma=0,~k=0$, Eq. (\ref{eq:metric}) reduces to the Reissner-Nordstr\"{o}m (RN) metric.
When $r$ is sufficiently large, the terms $1/r$, $1/r^2$, $1/r^6$ and $1/r^{14}$ become negligible, the metric approximates to
\begin{equation}
A(r) \approx 1 +\frac{8}{3} \alpha  (\alpha k+\pi )r^2.
\end{equation}
which is consistent with the AdS/dS metric, showing that the model reproduces the AdS/dS spacetime in the weak-field regime, thereby establishing its consistency with classical general relativity incorporating a cosmological constant. On the other hand, in the strong-field regions near the BH horizon and around the singularity, the $r^{-6}$ and $r^{-14}$ terms dominate the corrections.

\subsection{Horizon structure} 

Compared with the standard Reissner-Nordstr\"{o}m-AdS/dS (RN-AdS/dS) BH, the solution obtained here contains higher-order correction terms jointly dominated by nonlinear electrodynamic parameters $\alpha$, $\gamma$, and the modified gravity parameter $k$. These correction terms become significant at extremely small BH scales and are expected to substantially alter the number, positions, and properties of the horizons. We will investigate the evolutionary behavior of these horizon structures across different parameter spaces using both numerical and analytical methods.

The metric function may possess multiple zeros under certain parameters, endowing the BH with a complex structure. Analysis of the parameters shows how they distinctly affect the horizon configuration: increasing the mass $M$ enhances the negativity of the term $-\frac{2M}{r}$, causing the outer horizon to move outward; increasing the charge $Q$ makes $A(r)$ more positive primarily through the term $\frac{Q^2}{r^2}$, thus shrinking the outer horizon while expanding the inner horizon and approaching an extremal BH state. The parameter $\alpha$ influences the metric through both the effective cosmological constant term $\frac{8}{3} \alpha (\alpha k + \pi) r^2$ and the higher-order correction $\frac{4 \alpha \gamma k Q^4}{5 r^6}$; increasing $\alpha$ generally makes $A(r)$ more positive, potentially suppressing inner horizon formation while modifying the asymptotic spacetime structure. The NLED parameter $\gamma$ governs the strength of nonlinear electrodynamic effects: when $\gamma > 0$, the terms $-\frac{2 \pi \gamma Q^4}{5 r^6}$ and $-\frac{\gamma^2 k Q^8}{26 r^{14}}$ are negative, tending to decrease $A(r)$, which may eliminate existing inner horizons or generate new ones; when $\gamma < 0$, the signs term $-\frac{2 \pi \gamma Q^4}{5 r^6}$ reverse, producing opposite behavior. The coupling parameter $k$ of $f(R,T)$ amplifies matter-geometry interactions: when $k > 0$, competitive contributions emerge in terms of $r^{-6}$ and $r^{-14}$, with the net effect depending on the relative magnitudes of $\alpha$ and $\gamma$; when $k < 0$, the overall behavior undergoes reversal. The presence of the effective cosmological constant term $\frac{8}{3} \alpha (\alpha k + \pi) r^2$ significantly influences the asymptotic structure, introducing additional cosmological horizons in the appropriate parameter regimes. 

\begin{figure}[t]
    \centering
    \begin{minipage}[t]{0.45\textwidth}
        \centering
        \includegraphics[width=\textwidth]{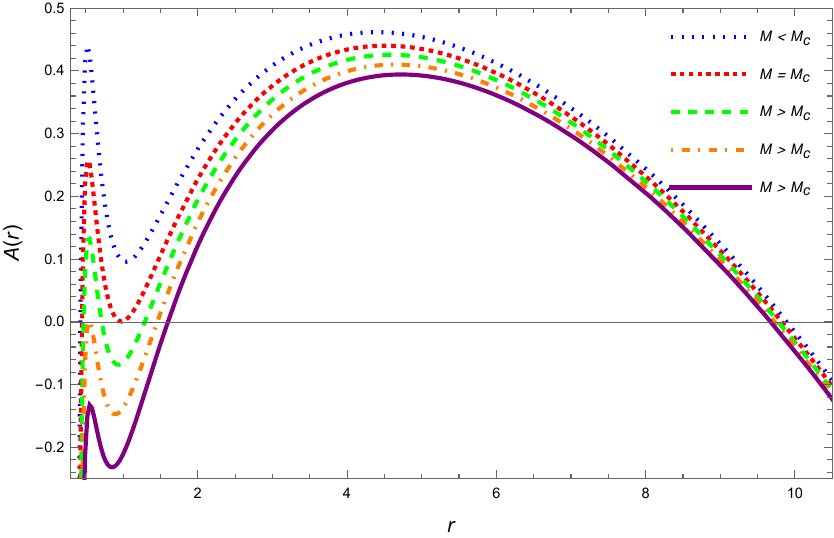}
        \caption{Plot of the BH metric function $A(r)$ with parameters $Q=1$, $\alpha=-0.001$, $\gamma=0.01$ and $k=0.02$.}
        \label{A-r,M.pdf}
    \end{minipage}
    \hfill
    \begin{minipage}[t]{0.45\textwidth}
        \centering
        \includegraphics[width=\textwidth]{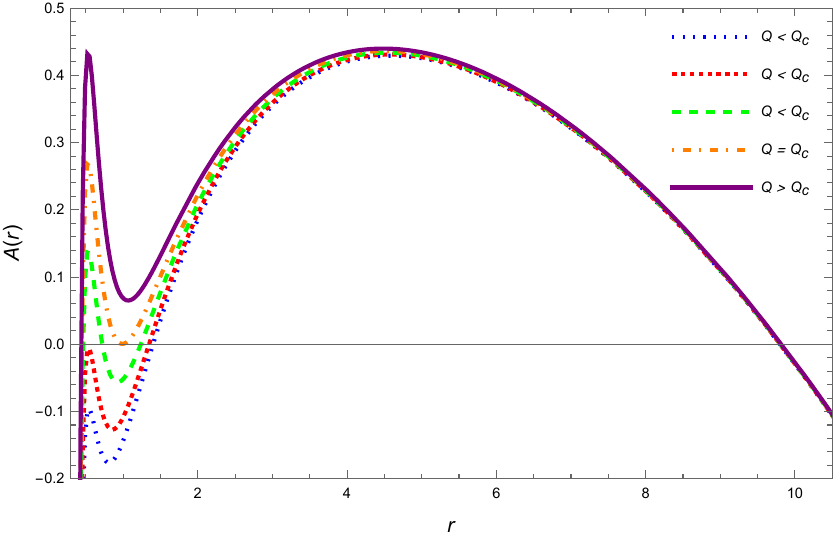}
        \caption{Plot of the BH metric function $A(r)$ with parameters $M=1$, $\alpha=-0.001$, $\gamma=0.01$ and $k=0.02$.}
        \label{A-r,Q.pdf}
    \end{minipage}
\end{figure}
We plot the metric function \eqref{eq:metric} for different values of $M$ and $Q$ in specific parameter configurations, as shown in Figs. \ref{A-r,M.pdf} and \ref{A-r,Q.pdf}. It can be observed that the influence of $M$ and $Q$ on the horizon structure shares common characteristics: both exhibit critical values ($M_c$ and $Q_c$) at which the BH develops three horizons: the innermost inner horizon, the intermediate event horizon, and the outermost cosmological horizon.
Let us now focus on the effect of $M$ on the horizon structure. When $M < M_c$, the BH possesses only the outer cosmological horizon and the inner event horizon. When $M > M_c$ and $M$ are sufficiently close to $M_c$, four horizons may appear, as illustrated by the green curve in the figures. As $M$ gradually increases further, the BH sequentially exhibits configurations with three and then two horizons.
The influence of $Q$ on the horizon structure is similar to that of $M$, but with opposite trends in variation.

The Kretschmann scalar, defined as the full contraction of the Riemann tensor with itself, \(K = R_{\mu\nu\rho\sigma} R^{\mu\nu\rho\sigma}\), serves as a curvature invariant that characterizes the true strength of the gravitational field. Its value remains finite at coordinate singularities (such as event horizons) and diverges only at physical singularities. For the black hole solution under consideration, we obtain the explicit expression
\begin{equation}
\begin{aligned}
K &= \frac{32 \gamma^2 Q^8 \left(11778 \alpha^2 k^2 r^4 + k \left(-13078 \pi \alpha r^4 + 750 M r - 1075 Q^2\right) + 3107 \pi^2 r^4\right)}{325 r^{20}} \\
&+ \frac{8 \left(2 r^2 \left(32 \alpha^2 r^6 (\alpha k + \pi)^2 + 9 M^2\right) - 36 M Q^2 r + 21 Q^4\right)}{3 r^8} - \frac{3664 \gamma^3 k Q^{12} (2 \alpha k - \pi)}{13 r^{24}} \\
&- \frac{64 \gamma Q^4 (\pi - 2 \alpha k) \left(20 \alpha r^4 (\alpha k + \pi) - 42 M r + 57 Q^2\right)}{15 r^{12}} + \frac{11222 \gamma^4 k^2 Q^{16}}{169 r^{32}}
\end{aligned}
\end{equation}
In summary, the Kretschmann scalar diverges when $r\to 0$, signaling a genuine curvature singularity at the center. As $r\to\infty$, it tends to a constant $\frac{64}{3}\alpha^2(\alpha k+\pi)^2$, indicating that the spacetime is not asymptotically flat due to the presence of an effective cosmological constant. At the event horizon, $K$ remains finite, confirming that the horizon is merely a coordinate singularity and not a physical one.

\section{Analysis of Thermodynamic Quantities}
\label{4}
In this section, we investigate the thermodynamic properties of static spherically symmetric magnetically charged AdS/dS black holes. We first recall the first law of black hole thermodynamics, then compute the Hawking temperature, entropy, Helmholtz free energy, internal energy, and heat capacity as functions of the event horizon radius, and subsequently discuss the stability and phase transitions of the black holes.

For a static spherically symmetric black hole, the first law of thermodynamics can be written as
\begin{equation}
dM = T dS + \Phi dQ,
\end{equation}
where $M$ is the black hole mass, $T$ is the Hawking temperature, $S$ is the black hole entropy, $\Phi$ is the electric potential, and $Q$ is the electric charge. In this work, we consider the magnetically charged case, so the potential term corresponds to the contribution of the magnetic charge.

The mass $M$, as a function of the event horizon radius $r_{\mathrm{h}}$ and magnetic charge $Q$, can be solved from the event horizon condition $A(r_{\mathrm{h}})=0$, yielding the expression
\begin{equation}
M = \frac{Q^2}{2r_{\mathrm{h}}} + \frac{1}{6} r_{\mathrm{h}} \left(3 + 8\pi \alpha r_{\mathrm{h}}^2 + 8k \alpha^2 r_{\mathrm{h}}^2\right) - \frac{Q^4 (\pi - 2k\alpha)\gamma}{5 r_{\mathrm{h}}^5} - \frac{k Q^8 \gamma^2}{52 r_{\mathrm{h}}^{13}}.
\end{equation}

The entropy of the black hole satisfies the Bekenstein area law, i.e.,
\begin{equation}
S = \pi r_{\mathrm{h}}^2.
\end{equation}

\begin{figure*}[ht]
    \centering
    \begin{minipage}[t]{0.48\textwidth}
        \centering
        \includegraphics[width=\linewidth]{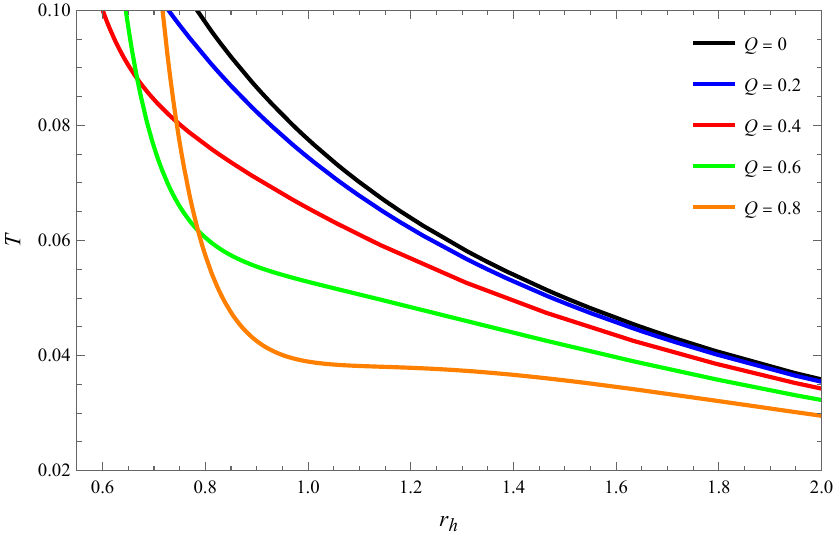}
    \end{minipage}
    \hfill
    \begin{minipage}[t]{0.48\textwidth}
        \centering
        \includegraphics[width=\linewidth]{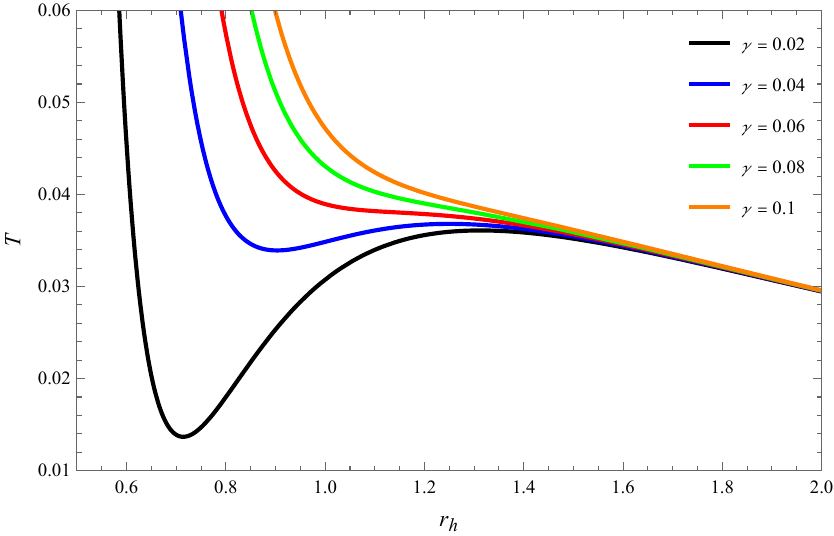}
    \end{minipage}
    \caption{The left panel shows the Hawking temperature $T$ as a function of the event horizon radius $r_{\mathrm{h}}$ for different magnetic charges $Q$, with fixed parameters  $\alpha = -0.001$, $k = 0.6$, $\gamma = 0.06$; the right panel shows $T$ as a function of $r_{\mathrm{h}}$ for different NLED parameters $\gamma$, with fixed parameters $\alpha = -0.001$, $Q = 0.8$, $k = 0.6$.}
    \label{T}
\end{figure*}

The Hawking temperature is a fundamental concept in black hole thermodynamics, describing the temperature at the black hole horizon, proportional to its surface gravity. For a stationary black hole, the Hawking temperature can be obtained from the partial derivative of the mass with respect to the entropy,
\begin{equation}
T = \left( \frac{\partial M}{\partial S} \right)_Q = \frac{(\partial M / \partial r_{\mathrm{h}})_Q}{(\partial S / \partial r_{\mathrm{h}})_Q}.
\end{equation}
Substituting the mass function yields the explicit expression for the Hawking temperature,
\begin{equation}
T = \frac{-2 Q^2 r_{\mathrm{h}}^{12} + 2 r_{\mathrm{h}}^{14} \bigl(1 + 8\pi \alpha r_{\mathrm{h}}^2 + 8k \alpha^2 r_{\mathrm{h}}^2\bigr) + 4 Q^4 r_{\mathrm{h}}^8 (\pi - 2k\alpha)\gamma + k Q^8 \gamma^2}{8\pi r_{\mathrm{h}}^{15}}.
\end{equation}
Fig. \ref{T} displays the evolution of the Hawking temperature as a function of the event horizon radius. The left panel shows that, for fixed NLED parameters and gravitational coupling parameters, the Hawking temperature exhibits a monotonic behavior with respect to the event horizon radius. Meanwhile, when the horizon radius is sufficiently large, the Hawking temperature corresponding to the same horizon radius decreases significantly as the magnetic charge $Q$ increases. The right panel illustrates the effect of the NLED parameter $\gamma$: as $\gamma$ increases, the Hawking temperature for a given horizon radius gradually rises in the small horizon radius region. When $\gamma$ takes smaller values, the overall trend of the temperature curve undergoes a significant change, exhibiting distinct maxima and minima.

\begin{figure*}[ht]
    \centering
    \begin{minipage}[t]{0.48\textwidth}
        \centering
        \includegraphics[width=\linewidth]{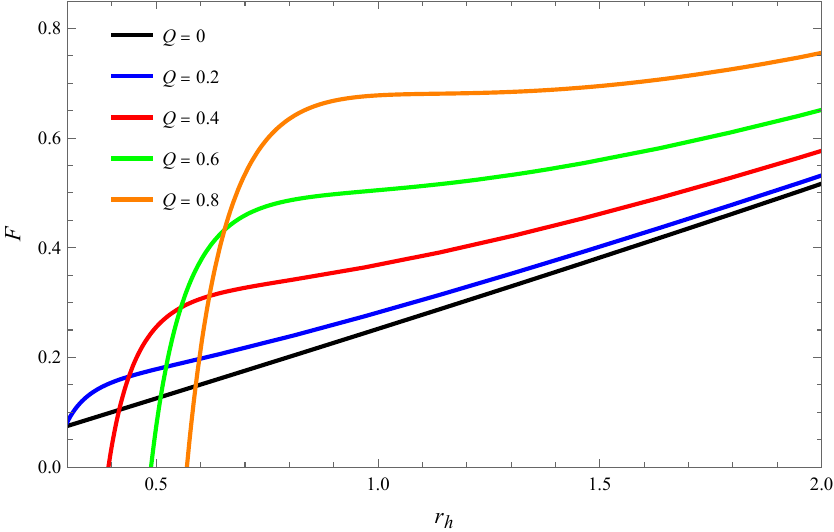}
    \end{minipage}
    \hfill
    \begin{minipage}[t]{0.48\textwidth}
        \centering
        \includegraphics[width=\linewidth]{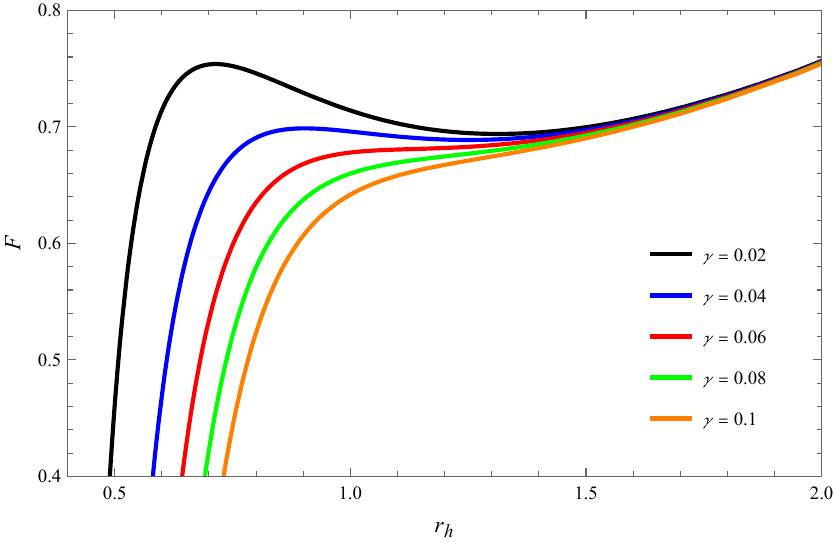}
    \end{minipage}
    \caption{The left panel shows the Helmholtz free energy $F$ as a function of the event horizon radius $r_{\mathrm{h}}$ for different magnetic charges $Q$, with fixed parameters $\alpha = -0.001$, $k = 0.6$, $\gamma = 0.06$; the right panel shows $F$ as a function of $r_{\mathrm{h}}$ for different NLED parameters $\gamma$, with fixed parameters $\alpha = -0.001$, $Q = 0.8$, $k = 0.6$.}
    \label{F}
\end{figure*}

The Helmholtz free energy is a thermodynamic potential that measures the ability of a system to perform work under constant temperature conditions. For a black hole, it reflects the thermodynamic stability at fixed temperature. It is defined as $F = M - T S$, and substituting $M$ and $T$ yields
\begin{equation}
F = \frac{3Q^2}{4 r_{\mathrm{h}}} + \frac{r_{\mathrm{h}}}{4} - \frac{2}{3} r_{\mathrm{h}}^3 \alpha (\pi + k\alpha) - \frac{7 Q^4 (\pi - 2k\alpha)\gamma}{10 r_{\mathrm{h}}^5} - \frac{15 k Q^8 \gamma^2}{104 r_{\mathrm{h}}^{13}}.
\end{equation}
Fig. \ref{F} presents the evolution of the Helmholtz free energy as a function of the event horizon radius. The left panel shows that for all magnetic charge values, the Helmholtz free energy increases monotonically with the horizon radius. Furthermore, as the magnetic charge $Q$ increases, the Helmholtz free energy corresponding to the same horizon radius rises significantly when the horizon radius is sufficiently large. In the right panel, as the NLED parameter $\gamma$ increases, the Helmholtz free energy for a given horizon radius gradually decreases, an effect that is particularly pronounced in the small horizon radius region. When $\gamma$ takes smaller values, the overall trend of the curve changes markedly, exhibiting clear maxima and minima.

\begin{figure*}[ht]
    \centering
    \begin{minipage}[t]{0.48\textwidth}
        \centering
        \includegraphics[width=\linewidth]{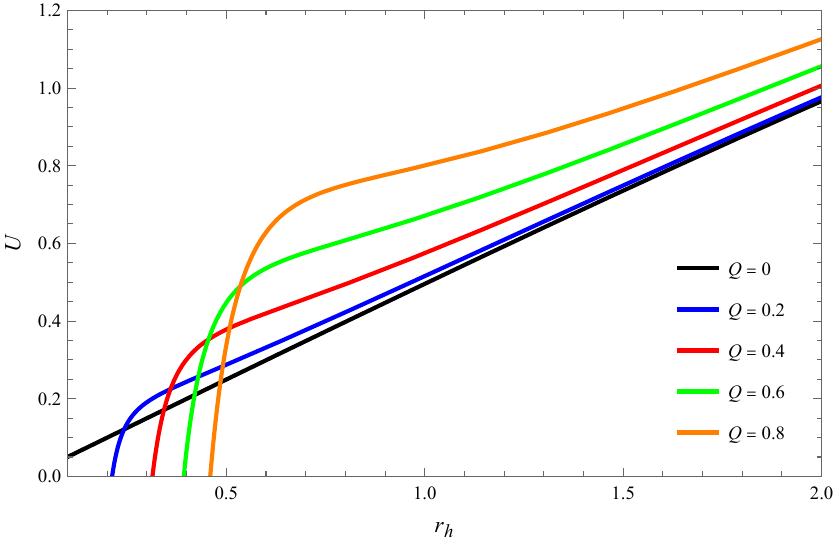}
    \end{minipage}
    \hfill
    \begin{minipage}[t]{0.48\textwidth}
        \centering
        \includegraphics[width=\linewidth]{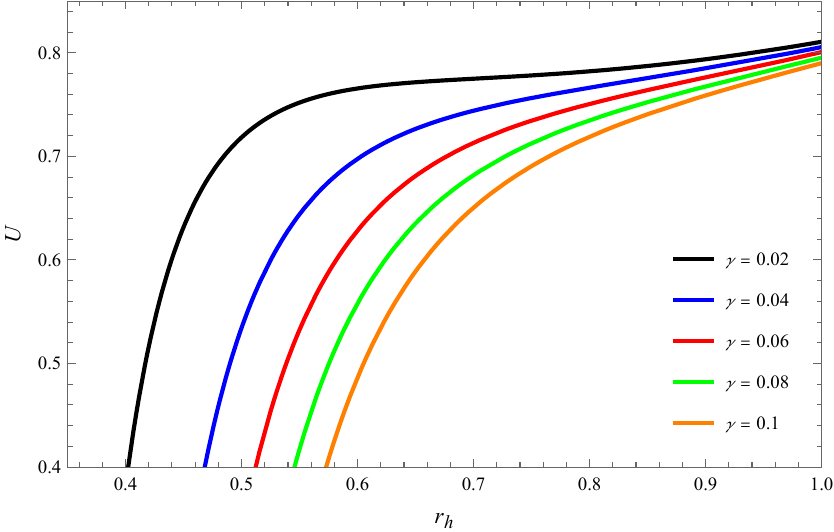}
    \end{minipage}
    \caption{The left panel shows the internal energy $U$ as a function of the event horizon radius $r_{\mathrm{h}}$ for different magnetic charges $Q$, with fixed parameters $\alpha = -0.001$, $k = 0.6$, $\gamma = 0.06$; the right panel shows $U$ as a function of $r_{\mathrm{h}}$ for different NLED parameters $\gamma$, with fixed parameters $\alpha = -0.001$, $Q = 0.8$, $k = 0.6$.}
    \label{U}
\end{figure*}

Internal energy is a measure of the energy contained within a system. From the relationship between the Helmholtz free energy, entropy, and temperature, we have
\begin{equation}
U = F + T S = M,
\end{equation}
meaning the internal energy equals the mass, with the explicit expression
\begin{equation}
U = \frac{Q^2}{2r_{\mathrm{h}}} + \frac{1}{6} r_{\mathrm{h}} \left(3 + 8\pi \alpha r_{\mathrm{h}}^2 + 8k \alpha^2 r_{\mathrm{h}}^2\right) - \frac{Q^4 (\pi - 2k\alpha)\gamma}{5 r_{\mathrm{h}}^5} - \frac{k Q^8 \gamma^2}{52 r_{\mathrm{h}}^{13}}.
\end{equation}
Fig. \ref{U} shows the evolution of the black hole internal energy as a function of the event horizon radius. As can be seen, for the given values of the magnetic charge and NLED parameters, the internal energy increases monotonically with $r_{\mathrm{h}}$. The left panel shows that for a sufficiently large horizon radius, the internal energy of the black hole becomes higher with a larger magnetic charge $Q$ at the same horizon radius. This is because the magnetic charge contributes additional electromagnetic energy to the black hole system, and this behavior is consistent with the physical picture of the standard RN black hole. Notably, in the small horizon radius region, the differences among the internal energy curves for different magnetic charges are significantly amplified, reflecting the strong-field effects introduced by the coupling between $f(R,T)$ gravity and NLED. The right panel illustrates the effect of the NLED parameter $\gamma$: as $\gamma$ increases, the internal energy for a given horizon radius gradually decreases, indicating that higher-order self-interactions in NLED weaken the energy contribution of the electromagnetic field.

\begin{figure*}[ht]
    \centering
    \begin{minipage}[t]{0.48\textwidth}
        \centering
        \includegraphics[width=\linewidth]{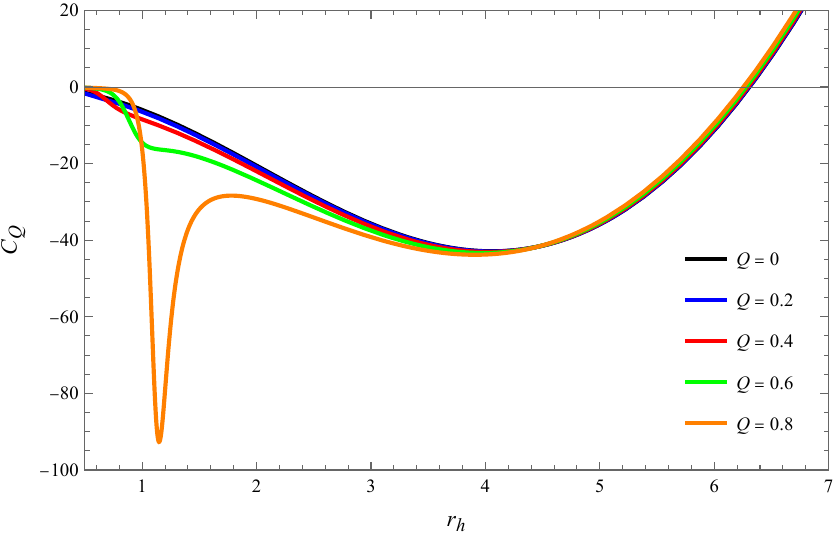}
    \end{minipage}
    \hfill
    \begin{minipage}[t]{0.48\textwidth}
        \centering
        \includegraphics[width=\linewidth]{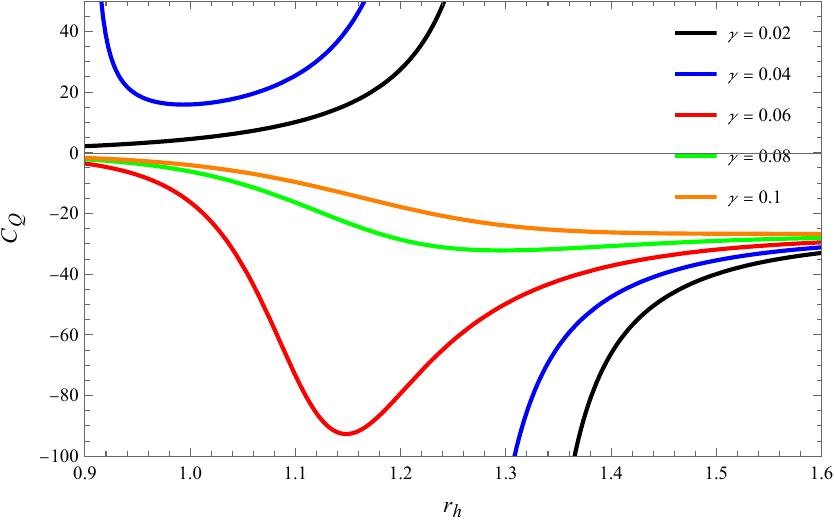}
    \end{minipage}
    \caption{The left panel shows the heat capacity $C_Q$ as a function of the event horizon radius $r_{\mathrm{h}}$ for different magnetic charges $Q$, with fixed parameters $\alpha = -0.001$, $k = 0.6$, $\gamma = 0.06$; the right panel shows $C_Q$ as a function of $r_{\mathrm{h}}$ for different NLED parameters $\gamma$, with fixed parameters $\alpha = -0.001$, $Q = 0.8$, $k = 0.6$.}
    \label{C}
\end{figure*}

The heat capacity is an important indicator of black hole thermodynamic stability, reflecting the response of the system's temperature to heat absorption. A positive heat capacity indicates stability, while a negative value signals instability. It is defined as
\begin{equation}
C_Q = T \left( \frac{\partial S}{\partial T} \right)_Q = T \frac{dS/dr_{\mathrm{h}}}{dT/dr_{\mathrm{h}}}.
\end{equation}
Using the above expressions, one obtains
\begin{equation}
C_Q = \frac{2\pi r_{\mathrm{h}}^2 \Bigl(-2 Q^2 r_{\mathrm{h}}^{12} + 2 r_{\mathrm{h}}^{14}\bigl(1 + 8\pi \alpha r_{\mathrm{h}}^2 + 8k \alpha^2 r_{\mathrm{h}}^2\bigr) + 4 Q^4 r_{\mathrm{h}}^8 (\pi - 2k\alpha)\gamma + k Q^8 \gamma^2\Bigr)}{6 Q^2 r_{\mathrm{h}}^{12} + 2 r_{\mathrm{h}}^{14}\bigl(-1 + 8\pi \alpha r_{\mathrm{h}}^2 + 8k \alpha^2 r_{\mathrm{h}}^2\bigr) - 28 Q^4 r_{\mathrm{h}}^8 (\pi - 2k\alpha)\gamma - 15 k Q^8 \gamma^2}.
\end{equation}
Fig. \ref{C} presents the evolution of the black hole heat capacity as a function of the event horizon radius. The divergence points of the heat capacity correspond to the critical points of second-order thermodynamic phase transitions and are a key feature of black hole thermodynamics. The left panel shows that the heat capacity exhibits a typical divergent jump behavior with respect to the horizon radius: in the small horizon radius region, the heat capacity is negative, indicating a locally unstable state. As $r_{\mathrm{h}}$ increases, the heat capacity undergoes an infinite jump from negative to positive at a critical radius, after which it remains positive, and the black hole enters a locally stable state. Notably, as the magnetic charge $Q$ increases, the heat capacity develops a trough before the jump, followed by a rapid recovery. In the right panel, as the NLED parameter $\gamma$ increases, the heat capacity transitions from a stable–unstable phase change regime to a unstable regime, and the phase transition critical point shifts toward smaller horizon radii, causing the black hole to switch from stable to unstable at smaller scales. Furthermore, compared with the standard RN-AdS black hole, the coupled corrections from $f(R,T)$ gravity and NLED significantly alter the location of the phase transition critical point and the quantitative behavior of the heat capacity, providing a theoretical criterion for distinguishing modified gravity from general relativity through thermodynamic observations.
\FloatBarrier

\section{The Effective metric and the black hole shadow}
\label{5}

\subsection{The effective potential}

In vacuum linear electrodynamics (Maxwell theory), electromagnetic waves are linear and their solutions are independent. A photon propagates without interacting with other photons or background electromagnetic fields. Therefore, it perceives the structure of the background spacetime itself, and its propagation path is described by the null geodesics determined by the background spacetime metric $g_{\mu\nu}$.

In nonlinear electrodynamics, the situation changes. The Lagrangian includes higher-order terms (such as $F^2, F^3, \dots$). This implies that the electromagnetic field is self-interacting: the propagation of one photon is influenced by other photons or background electromagnetic fields. In this case, light rays propagate along the geodesics of an effective metric which is given by metric\cite{Novello:1999pg,Toshmatov:2021fgm}
\begin{equation}
    \label{eq:effective metric}
    g_{\mathrm{eff}}^{\mu\nu} =\mathcal{L}_F g^{\mu\nu}-\mathcal{L}_{FF} F_\sigma^\mu F^{\sigma\nu}.
\end{equation}
For the magnetically charged BH discussed here, let the line element be
\begin{equation}
     ds^2 = -f(r) dt^2 +g(r)dr^2 + h(r) (d\theta^2 + \sin^2\theta d\phi^2),
\end{equation}
According to Eq. (\ref{eq:effective metric}), we can derive
\begin{equation}
f(r)=\frac{A(r)}{\mathcal{L}_F},
\end{equation}
\begin{equation}
g(r)=\frac{1}{A(r)\mathcal{L}_F},
\end{equation}
\begin{equation}
h(r)=\frac{r^2}{\mathcal{L}_F + 2F\mathcal{L}_{FF}}.
\end{equation}
The metric $g_{\mathrm{eff}}^{\mu\nu}$ does not explicitly depend on the coordinates $t$ and $\phi$, therefore, this spacetime possesses two independent Killing vector fields. So for a null geodesic, both the energy $E$ and the angular momentum $L$ are conserved
\begin{equation}
\label{E}
E=-g_{\mathrm{eff}\,00}\frac{dt}{d\tau}=f(r)\frac{dt}{d\tau},
\end{equation}
\begin{equation}
\label{L}
L=g_{\mathrm{eff}\,33}\frac{d\phi}{d\tau}=h(r)\frac{dt}{d\tau},
\end{equation}
where $\tau$ is the affine parameter. Define the impact parameter $b=L/E$, accordingly, the equation satisfied by the null geodesic can be written as
\begin{equation}
\label{eq:geo0}
\left(\frac{dr}{d\phi}\right)^2=\frac{h(r)}{g(r)}\left[\frac{h(r)}{f(r)}\frac{1}{b^2}-1\right].
\end{equation}
We define $V_{\mathrm{eff}}(r) = f(r)/h(r)$ as the effective potential. Photons will orbit the BH in an unstable circular trajectory, forming a photon sphere if the effective potential satisfies the following conditions:
\begin{equation}
\frac{dV_{\mathrm{eff}}(r_{\rm ph})}{dr} = 0, 
\end{equation}
\begin{equation}
\frac{d^2V_{\mathrm{eff}}(r_{\rm ph})}{dr^2} \le 0,
\end{equation}
where $r_{\rm ph}$ is the radius of the photon sphere, and the corresponding $V_{\mathrm{eff}}(r)$ is at its maximum, with
\begin{equation}
b_{\rm ph} = \frac{1}{\sqrt{V_{\mathrm{eff}}(r_{\rm ph})}},
\end{equation}
where $b_{\rm ph}$ represents the critical impact parameter.

\begin{figure}[t]
    \centering
    \begin{minipage}[t]{0.45\textwidth}
        \centering
        \includegraphics[width=\textwidth]{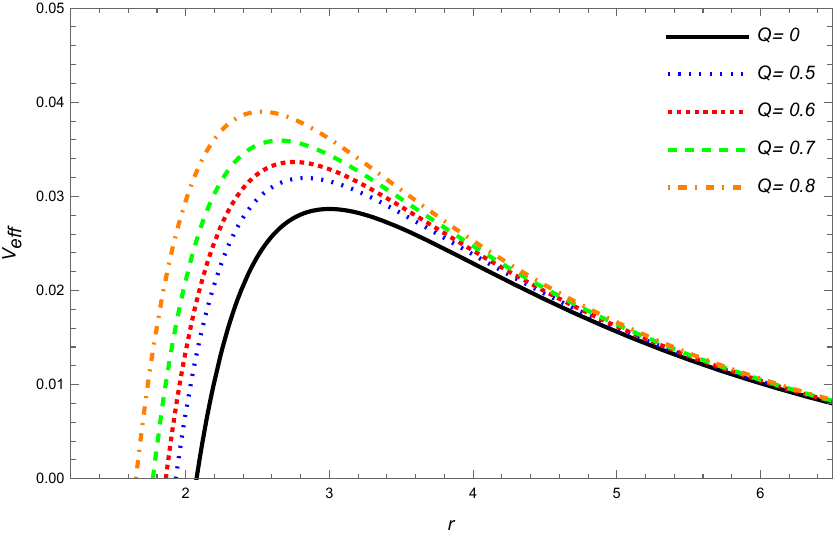}
        \caption{Plot of the effective potential function $V_{\mathrm{eff}}(r)$ with parameters $M=1$, $\alpha=-0.001$, $\gamma=0.06$, and $k=0.6$.}
        \label{Veff-r,Q.pdf}
    \end{minipage}
    \hfill
    \begin{minipage}[t]{0.45\textwidth}
        \centering
        \includegraphics[width=\textwidth]{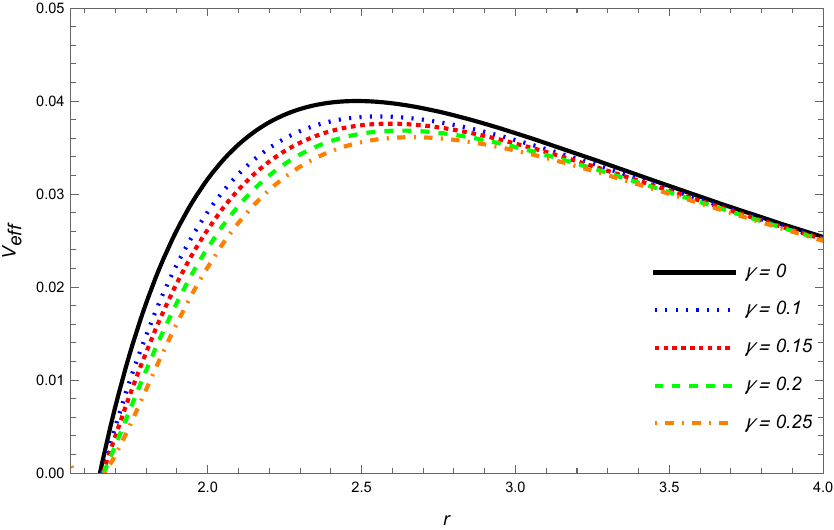}
        \caption{Plot of the effective potential function $V_{\mathrm{eff}}(r)$ with parameters $M=1$, $Q=0.8$, $\alpha=-0.001$, and $k=0.6$.}
        \label{Veff-r,Gamma.pdf}
    \end{minipage}
\end{figure}

Figure. \ref{Veff-r,Q.pdf} illustrates the influence of the magnetic charge $Q$ on the effective potential. The effective potential curves $V_{\mathrm{eff}}(r)$ show significant variations under different charge parameters $Q$. When $Q = 0$ (Schwarzschild BH), the effective potential exhibits a prominent maximum at $r = 3M$, corresponding to an unstable circular-photon orbit. As the charge $Q$ increases, the height of this potential barrier increases and its position shifts inward, leading to a decrease in the radius of BH shadow. When $Q$ is sufficiently large (e.g., $Q = 0.8$), the effective potential shows a tendency to diverge, suggesting that the photon sphere may cease to exist for sufficiently large magnetic charges.

Figure. \ref{Veff-r,Gamma.pdf} demonstrates the variation of effective potential with different values of $\gamma$. As the value of $\gamma$ increases, the height of the potential barrier decreases significantly, and the peak position shifts towards larger $r$ values, resulting in an increase in the radius of the BH shadow. This indicates that the parameter $\gamma$ has a substantial impact on the spacetime geometry, weakening the binding capacity of the effective potential in photons.

\subsection{The shadow}
To further discuss the characteristics of photon orbits, we substitute $u=1/r$ into Eq. (\ref{eq:geo0}) and rewrite it as follows
\begin{equation}
\label{eq:geo0re}
\left(\frac{du}{d\phi}\right)^2=\frac{h\left(\frac{1}{u}\right)u^4}{g\left(\frac{1}{u}\right)}\left[\frac{h\left(\frac{1}{u}\right)}{f\left(\frac{1}{u}\right)}\frac{1}{b^2}-1\right].
\end{equation}
For the case of light directly infalling into the BH, corresponding to an impact parameter $b<b_{\rm ph}$, the total change in azimuthal angle as the photon travels from infinity to the event horizon can be obtained by integrating over $u$
\begin{equation}
\Delta\phi=\int_{0}^{\frac{1}{r_h}} \frac{1}{\sqrt{\frac{h\left(\frac{1}{u}\right)u^4}{g\left(\frac{1}{u}\right)}\left[\frac{h\left(\frac{1}{u}\right)}{f\left(\frac{1}{u}\right)}\frac{1}{b^2}-1\right]}}  du,
\end{equation}
here $r_{\rm h}$ is the radius of the event horizon. If the photon does not fall into the event horizon, corresponding to an impact parameter $b>b_{\rm ph}$, the total azimuthal angle for the photon traveling from infinity to the point of closest approach and then back to infinity is
\begin{equation}
\Delta\phi=\int_{0}^{\frac{1}{r_{\rm min}}} \frac{2}{\sqrt{\frac{h\left(\frac{1}{u}\right)u^4}{g\left(\frac{1}{u}\right)}\left[\frac{h\left(\frac{1}{u}\right)}{f\left(\frac{1}{u}\right)}\frac{1}{b^2}-1\right]}}  du,
\end{equation}
where $r_{\rm min}$ is the minimum distance of the photon from the BH.
Like in \cite{Gralla:2019xty}, we can define the total orbit number $n=\Delta\phi/2\pi$ and use it to describe the photon's behavior: denoting the impact parameters as $b^-_2$ and $b^-_3$, corresponding to $n=0.75$ and $n=1.25$ for photons falling into the BH, respectively; and denoting the impact parameters for scattered photons as $b^+_2$ and $b^+_3$, respectively. Based on these agreements, we classify the impact parameter intervals as follows
\begin{itemize}
    \item Direct emission: $0 < n < \frac{3}{4}$, corresponding to $b \in (0, b^-_2) \cup (b^+_2, \infty)$,
    \item Lensed: $\frac{3}{4} < n < \frac{5}{4}$, corresponding to $b \in (b^-_2, b^-_3) \cup (b^+_3, b^+_2)$,
    \item Photon ring: $n > \frac{5}{4}$, corresponding to $b \in (b^-_3, b^+_3)$.
\end{itemize}

\begin{table}[ht]
\centering
\caption{Numerical values of various parameters for different magnetic charges with $M=1$, $k=0.6$, $\alpha=-0.001$ and $\gamma=0.06$.}
\label{rbQ}
\begin{tabular}{|c|c|c|c|c|c|c|c|}
\hline
\textbf{} & \textbf{$r_{\rm h}$} & \textbf{$r_{\rm ph}$} & \textbf{$b_{\rm ph}$} & \textbf{$b^-_2$} & \textbf{$b^-_3$} & \textbf{$b^+_3$} & \textbf{$b^+_2$} \\
\hline
$Q=0$ & 2.07481 & 3.000 & 5.90682 & 5.66734 & 5.89573 & 5.95364 & 7.47166 \\
\hline
$Q=0.6$ & 1.86097 & 2.75231 & 5.451 & 5.19833 & 5.4376 & 5.503 & 6.99725 \\
\hline
$Q=0.7$ & 1.77085 & 2.65197 & 5.27559 & 5.01509 & 5.26089 & 5.33036 & 6.82163 \\
\hline
$Q=0.8$ & 1.65323 & 2.52663 & 5.06359 & 4.79099 & 5.04687 & 5.12249 & 6.60411 \\
\hline
\end{tabular}
\end{table}

\begin{figure}[ht]
    \centering
    % First row of images
    \subfigure[$Q=0$]{
        \includegraphics[width=0.229\textwidth]{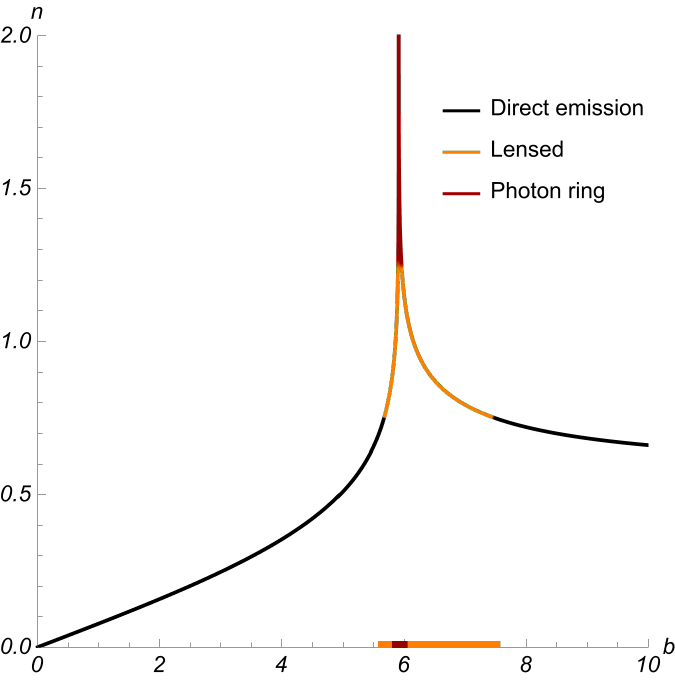}
        \label{n-Q=0.pdf}
    }
    \hfill
    \subfigure[$Q=0.6$]{
        \includegraphics[width=0.229\textwidth]{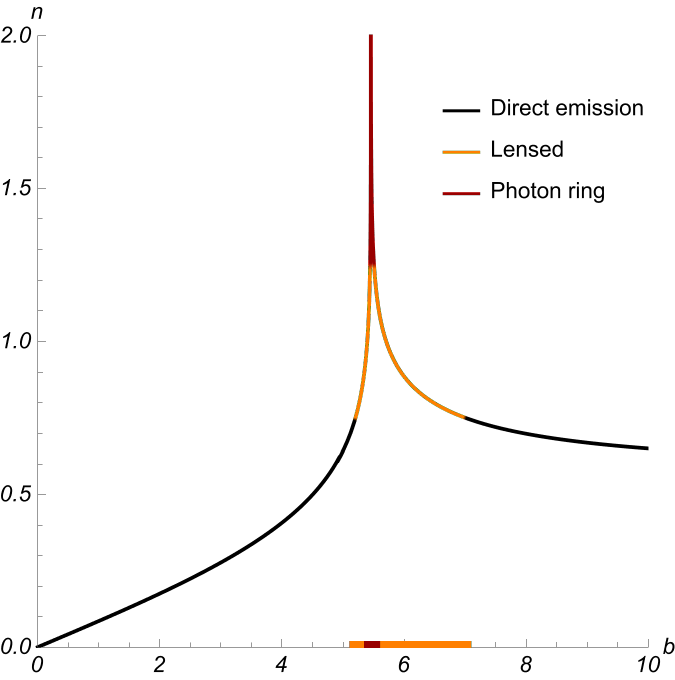}
        \label{n-Q=0.6.pdf}
    }
    \hfill
    \subfigure[$Q=0.7$]{
        \includegraphics[width=0.229\textwidth]{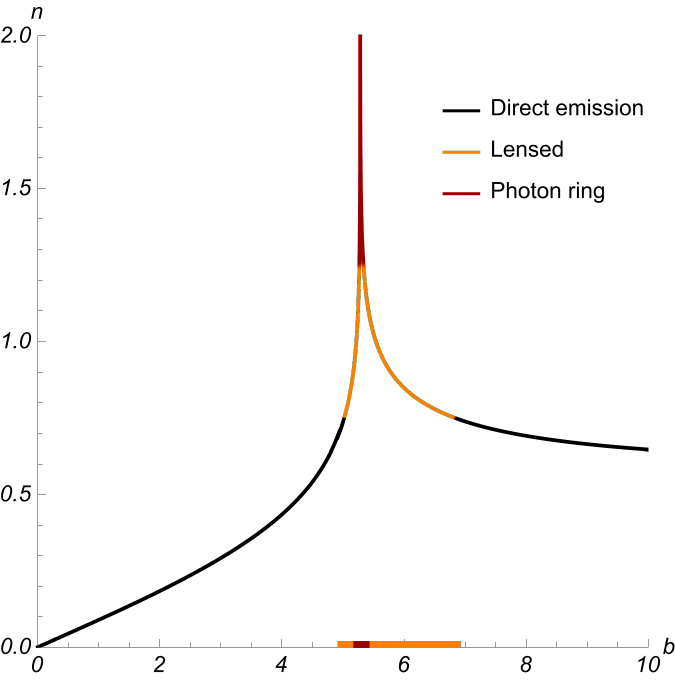}
        \label{n-Q=0.7.pdf}
    }
    \hfill
    \subfigure[$Q=0.8$]{
        \includegraphics[width=0.229\textwidth]{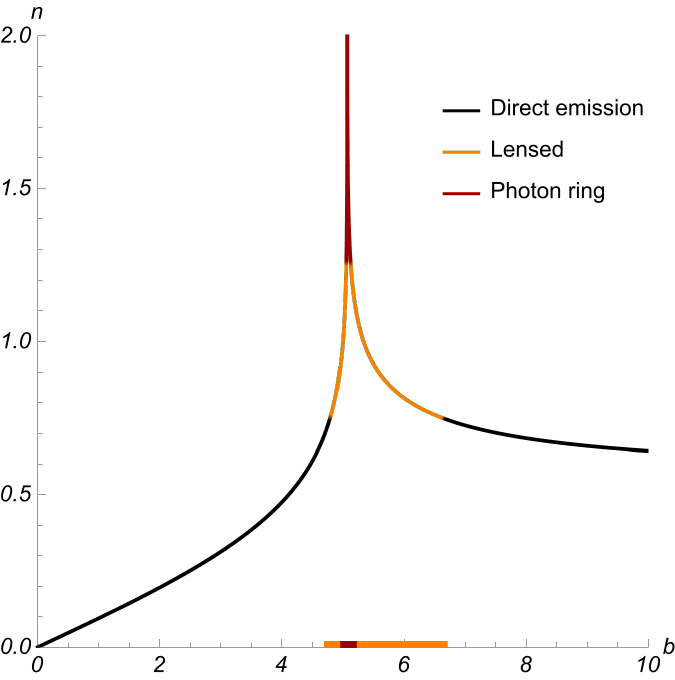}
        \label{n-Q=0.8.pdf}
    }
    
    \vspace{0.5cm} % Row spacing
    
    % Second row of images
    \subfigure[$Q=0$]{
        \includegraphics[width=0.229\textwidth]{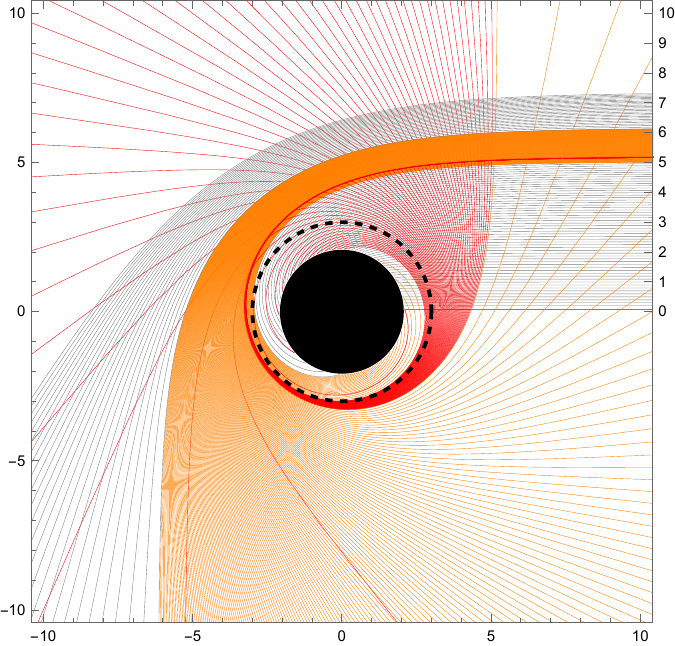}
        \label{geo-Q=0.pdf}
    }
    \hfill
    \subfigure[$Q=0.6$]{
        \includegraphics[width=0.229\textwidth]{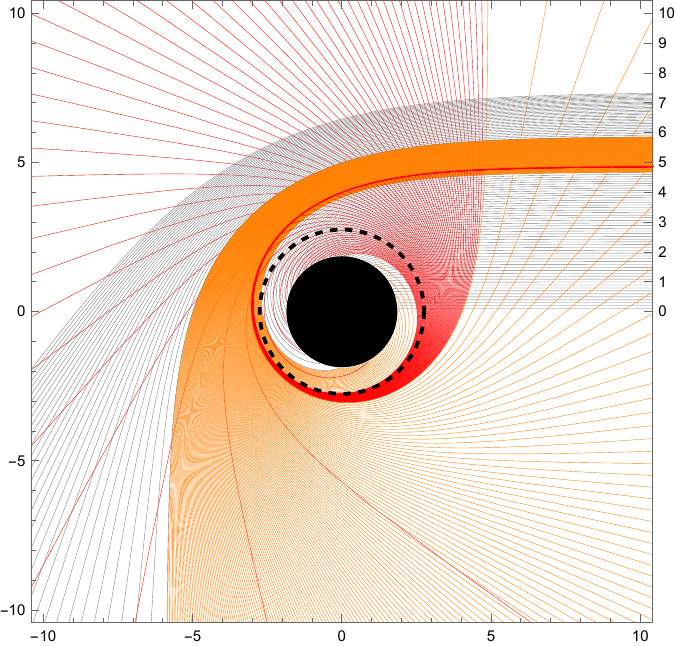}
        \label{geo-Q=0.6.pdf}
    }
    \hfill
    \subfigure[$Q=0.7$]{
        \includegraphics[width=0.229\textwidth]{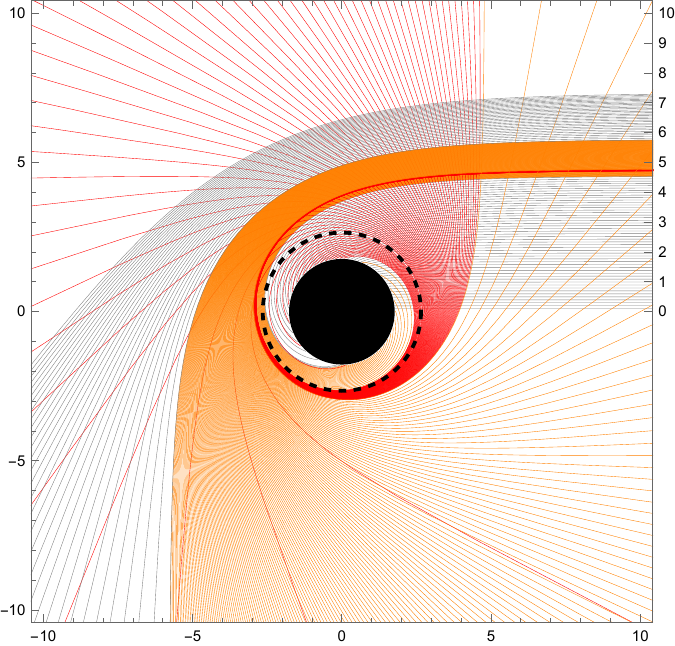}
        \label{geo-Q=0.7.pdf}
    }
    \hfill
    \subfigure[$Q=0.8$]{
        \includegraphics[width=0.229\textwidth]{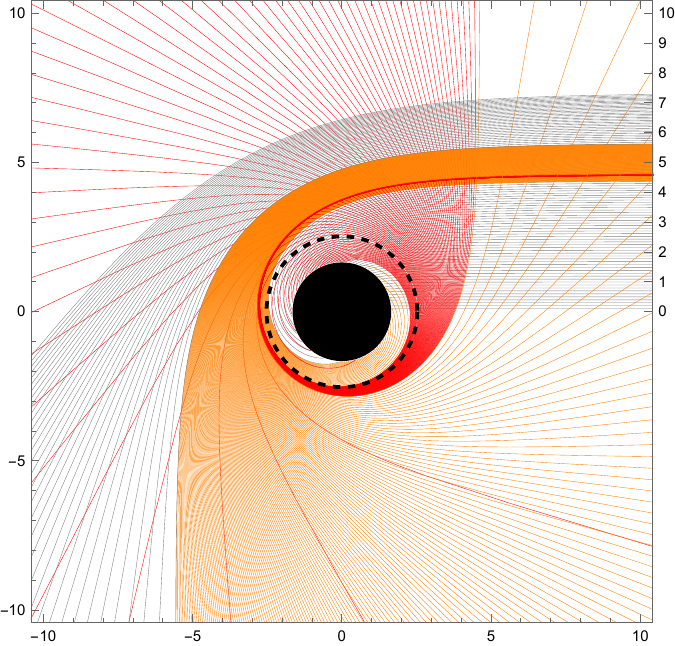}
        \label{geo-Q=0.8.pdf}
    }
    \caption{Photon geodesic structure described by the effective metric. The first row shows the relationship between the orbit number $n$ and the impact parameter $b$; the second row displays the photon geodesic curves with $M=1$, $k=0.6$, $\alpha=-0.001$ and $\gamma=0.06$.}
    \label{geoQ}
\end{figure}

The numerical results presented in Table. \ref{rbQ} and Figure. \ref{geoQ} reveal that as the magnetic charge increases, the event horizon radius and the photon sphere radius contract significantly, and the critical impact parameter $b_{\rm ph}$ decreases. Furthermore, a larger total orbit number $n$ is achieved with a smaller impact parameter.

\begin{table}[ht]
\centering
\caption{Numerical values of various parameters for different $\gamma$ with $M=1$, $Q=0.8$, $k=0.6$ and $\alpha=-0.001$.}
\label{rbgamma}
\begin{tabular}{|c|c|c|c|c|c|c|c|}
\hline
\textbf{} & \textbf{$r_{\rm h}$} & \textbf{$r_{\rm ph}$} & \textbf{$b_{\rm ph}$} & \textbf{$b^-_2$} & \textbf{$b^-_3$} & \textbf{$b^+_3$} & \textbf{$b^+_2$} \\
\hline
$\gamma=0$ & 1.64964 & 2.48489 & 4.99921 & 4.70732 & 4.97978 & 5.064 & 6.58884 \\
\hline
$\gamma=0.15$ & 1.65847 & 2.59142 & 5.15872 & 4.91977 & 5.14603 & 5.20925 & 6.6544 \\
\hline
$\gamma=0.2$ & 1.6613 & 2.62783 & 5.21051 & 4.99212 & 5.19987 & 5.25674 & 6.67604 \\
\hline
$\gamma=0.25$ & 1.66409 & 2.66412 & 5.26137 & 5.06434 & 5.2526 & 5.30362 & 6.69757 \\
\hline
\end{tabular}
\end{table}

\begin{figure}[htb]
    \centering
    % First row of images
    \subfigure[$\gamma=0$]{
        \includegraphics[width=0.229\textwidth]{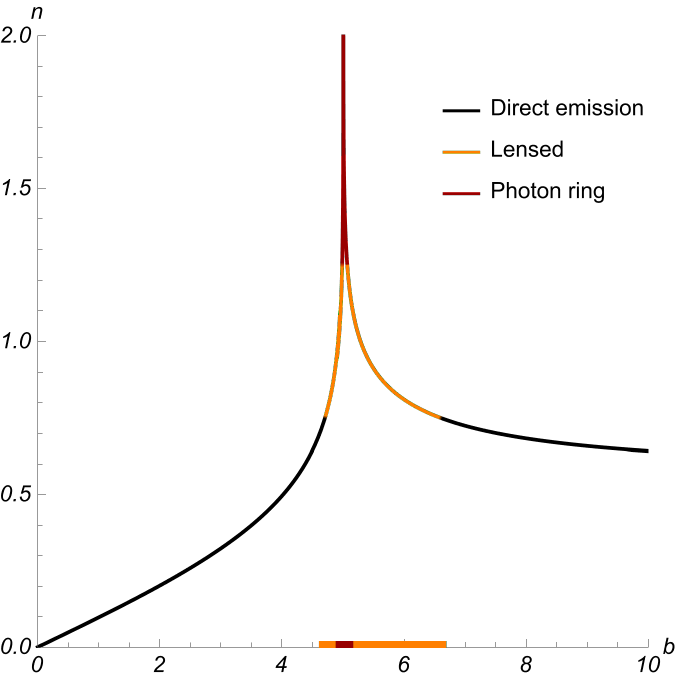}
        \label{n-gamma=0.pdf}
    }
    \hfill
    \subfigure[$\gamma=0.15$]{
        \includegraphics[width=0.229\textwidth]{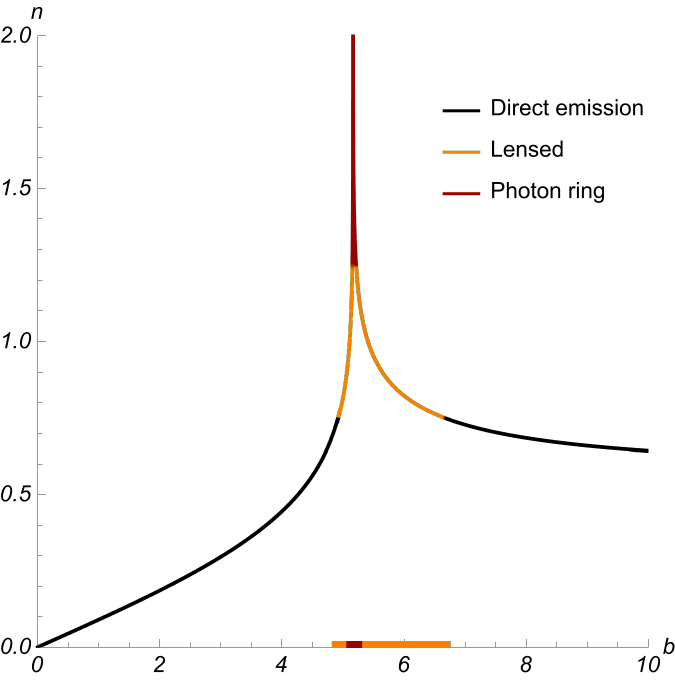}
        \label{n-gamma=0.15.pdf}
    }
    \hfill
    \subfigure[$\gamma=0.2$]{
        \includegraphics[width=0.229\textwidth]{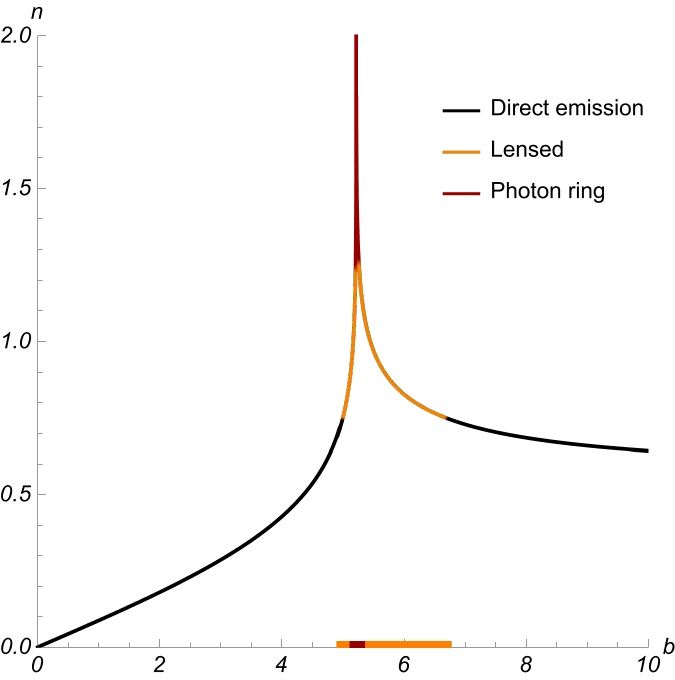}
        \label{n-gamma=0.2.pdf}
    }
    \hfill
    \subfigure[$\gamma=0.25$]{
        \includegraphics[width=0.229\textwidth]{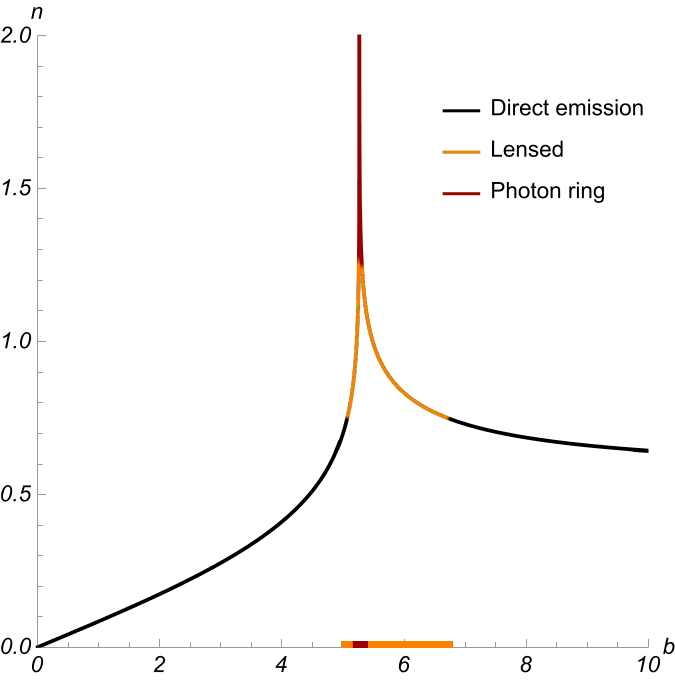}
        \label{n-gamma=0.25.pdf}
    }
    
    \vspace{0.5cm} % Row spacing
    
    % Second row of images
    \subfigure[$\gamma=0$]{
        \includegraphics[width=0.229\textwidth]{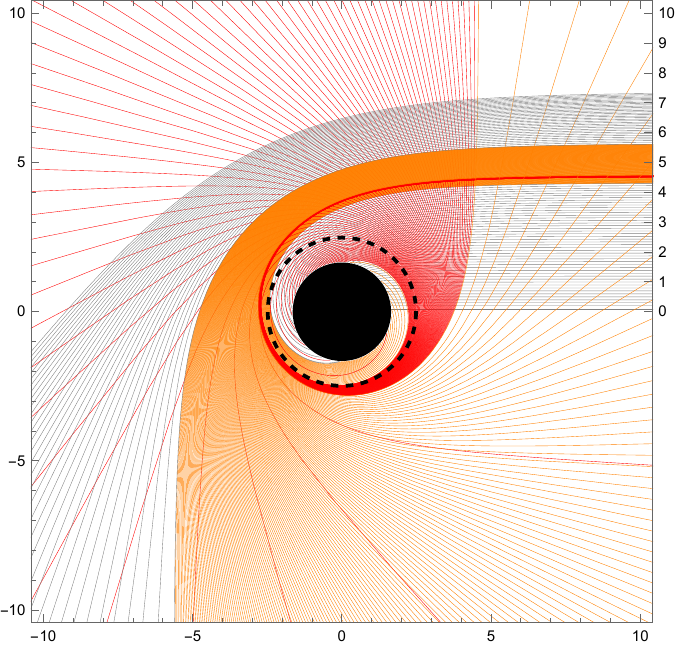}
        \label{geo-gamma=0.pdf}
    }
    \hfill
    \subfigure[$\gamma=0.15$]{
        \includegraphics[width=0.229\textwidth]{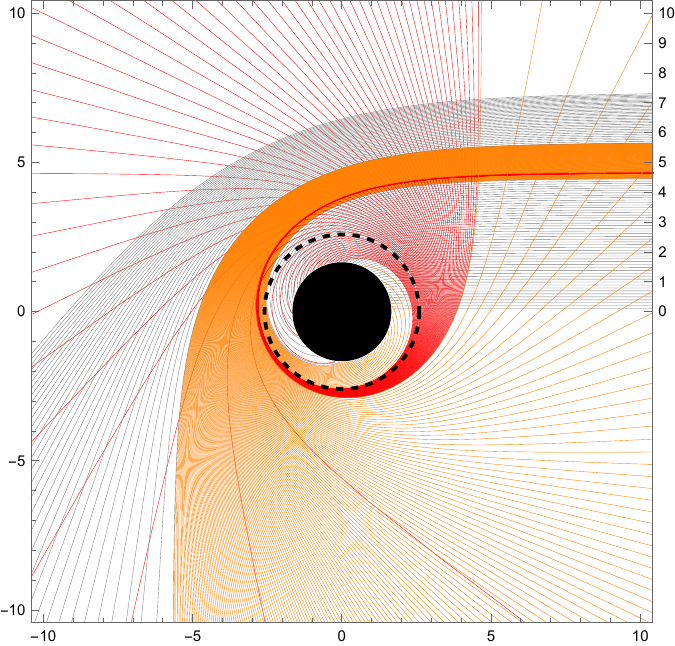}
        \label{geo-gamma=0.15.pdf}
    }
    \hfill
    \subfigure[$\gamma=0.2$]{
        \includegraphics[width=0.229\textwidth]{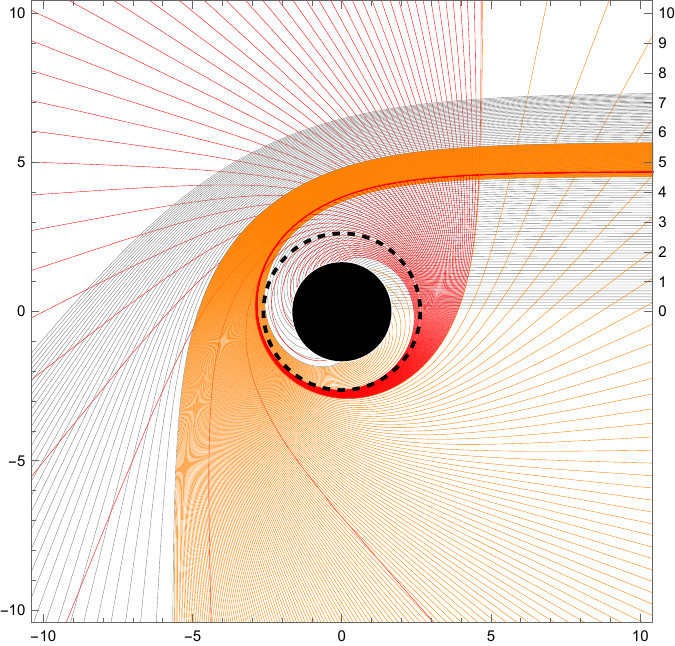}
        \label{geo-gamma=0.2.pdf}
    }
    \hfill
    \subfigure[$\gamma=0.25$]{
        \includegraphics[width=0.229\textwidth]{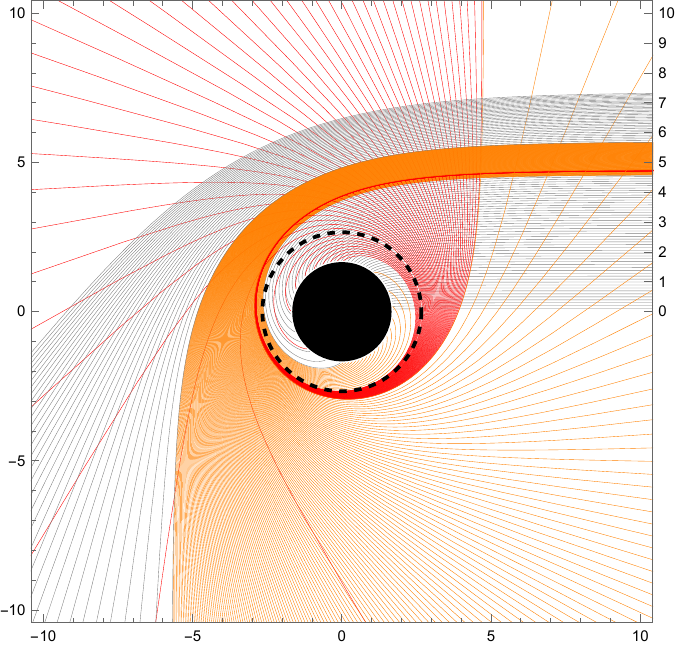}
        \label{geo-gamma=0.25.pdf}
    }
    \caption{Photon geodesic structure described by the effective metric. The first row shows the relationship between the orbit number $n$ and the impact parameter $b$; the second row displays the photon geodesic curves with $M=1$, $Q=0.8$, $k=0.6$ and $\alpha=-0.001$.}
    \label{geogamma}
\end{figure}

The computational results displayed in Table. \ref{rbgamma} and Figure. \ref{geogamma} indicate that as the coupling parameter $\gamma$ increases, the radius of the event horizon and the radius of the photon sphere expand, and the critical impact parameter $b_{\rm ph}$ increases. 

%Furthermore, achieving a specified value of $n$ requires a larger impact parameter.
\FloatBarrier
\subsection{Constraints on the Coupling Parameter}
\begin{figure*}[ht]
    \centering
    \begin{minipage}[t]{0.48\textwidth}
        \centering
        \includegraphics[width=\linewidth]{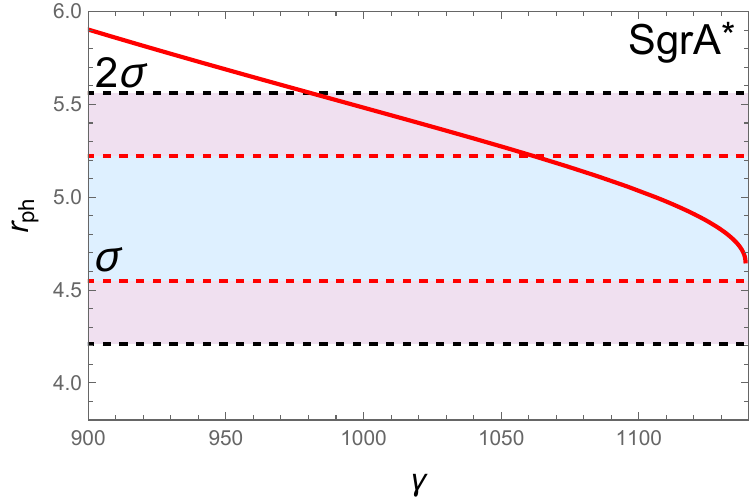}
    \end{minipage}
    \hfill
    \begin{minipage}[t]{0.48\textwidth}
        \centering
        \includegraphics[width=\linewidth]{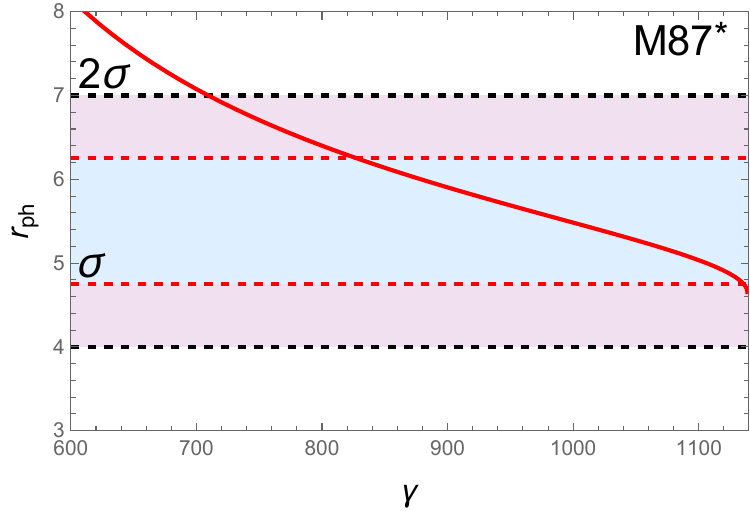}
    \end{minipage}
    \caption{The left panel shows the constraints on the model parameters from the observational data of Sgr A*, and the right panel shows those from M87*. The shadow radius of the black hole model is indicated by the solid red curve. The blue and pink shaded regions represent the $1\sigma$ and $2\sigma$ confidence level constraints on the shadow radius of Sgr A* and M87* from the EHT, respectively, while the gray region lies beyond the $2\sigma$ range.}
    \label{M87SgrA}
\end{figure*}

As shown in Fig. \ref{M87SgrA}, based on the observational data of the M87* and Sgr A* black hole shadows from the EHT collaboration, we constrain the model parameter $\gamma$. Under the fixed parameters $M=1$, $Q=0.5$, $\alpha=-0.001$, and $k=0.6$, and by matching the black hole shadow radius $r_{\mathrm{ph}}$ with the observational results, the constrained range for $\gamma$ is obtained as follows:
For the Sgr A* black hole, at the $1\sigma$ confidence level, the upper and lower bounds of $\gamma$ are $1062.22$ and $1139.23$, respectively; at the $2\sigma$ confidence level, the upper bound is $981.13$.
For the M87* black hole, at the $1\sigma$ confidence level, the upper and lower bounds of $\gamma$ are $827.759$ and $1135.79$, respectively; at the $2\sigma$ confidence level, the upper bound is $709.434$.

These results indicate that the theoretical model under consideration is compatible with the EHT observational data within the given parameter space, with the value of $\gamma$ constrained to the order of $10^3$, further validating the applicability of the model in tests of strong-field gravity.

\FloatBarrier

\section{Conclusions}	
\label{6}
Within the framework combining modified gravity with higher-order matter-geometry coupling and nonlinear electrodynamics containing quadratic terms, we have successfully derived an analytical solution for a static spherically symmetric magnetically charged (anti)de Sitter BH and systematically investigated its spacetime structure, thermodynamic properties, and photon orbits.

The fundamental theoretical advance of this research lies in the extension of matter-geometry coupling in $f(R,T)$ gravity from the commonly used linear form ($T$) to a quadratic form ($T^2$). This extension enables the energy-momentum tensor to source the spacetime curvature nonlinearly, inducing more profound modifications to spacetime geometry in strong-field regions. The derived metric function $A(r)$ clearly demonstrates this characteristic: beyond the standard Schwarzschild and Reissner-Nordstr\"{o}m terms, it incorporates higher-order correction terms arising from the coupling from $T^2$ and $\gamma F^2$ terms, along with an effective cosmological constant term. Although negligible in the asymptotic region, these terms dominate the strong-gravity core region, profoundly altering singularity behavior. The Kretschmann scalar analysis confirms that the curvature singularity at $r\to 0$ is genuine, while the spacetime approaches a constant curvature asymptotically, consistent with the presence of an effective cosmological constant.

Our analyses of the metric function reveal a more complex horizon structure compared to that of standard BHs. By adjusting the mass $M$, the magnetic charge $Q$, the gravitational coupling constant $k$, and the NLED parameters $\alpha$ and $\gamma$, the BH can exhibit various configurations including a single horizon, two horizons, three horizons, or even four horizons. The emergence of this multi-horizon structure is a direct consequence of the combined effects of modified gravity and nonlinear electromagnetic interactions.

The thermodynamic analysis demonstrates that the Hawking temperature, Helmholtz free energy, and internal energy exhibit monotonic behavior with respect to the horizon radius under fixed parameters, while the heat capacity reveals critical phase transition behavior. The divergence of the heat capacity signals second-order phase transitions, with the critical point shifting significantly due to the coupled corrections from $f(R,T)$ gravity and NLED, providing a potential observational signature to distinguish modified gravity from general relativity.

By introducing the concept of an effective metric, we have accurately described photon propagation paths in the context of nonlinear electrodynamics. We have found that the magnetic charge $Q$ and the NLED parameter $\gamma$ exert distinctly different influences on photon orbits: increasing the magnetic charge $Q$ reduces both the photon sphere radius $r_{\rm ph}$ and the critical impact parameter $b_{\rm ph}$, leading to a contraction of the BH shadow; conversely, increasing the NLED parameter $\gamma$ produces the opposite effect, significantly increasing both the photon sphere radius and the shadow size. This indicates that the nonlinear electromagnetic effects represented by $\gamma$ generate an additional repulsive interaction, weakening the gravitational binding of photons by the BH and requiring photons to orbit at greater distances to form stable circular trajectories. This phenomenon represents one of the key distinguishing features of the model compared to predictions of standard general relativity. By comparing the theoretical shadow radii with the EHT observational data for M87* and Sgr A*, we have placed constraints on the model parameter $\gamma$, which is found to be of order $10^3$ within the $1\sigma$ and $2\sigma$ confidence levels, demonstrating the consistency of the model with current strong-field gravity observations.

Finally, the present framework can be extended to rotating black holes, exotic configurations such as black-bounce solutions, and gravitational wave signatures, offering promising directions for further testing modified gravity in strong-field regimes.

\begin{acknowledgments}
This study is supported in part by National Natural Science Foundation of China (Grant No. 12333008).
\end{acknowledgments}

\bibliographystyle{ieeetr}
\bibliography{reff}
\end{document}